\documentclass[conference]{IEEEtran}

\usepackage{color}
\usepackage{graphicx}
\usepackage{amsfonts}
\usepackage{amsmath}
\usepackage{psfrag}
\usepackage{subfigure}
\usepackage{cite}
\usepackage{algorithmic}
\usepackage{amsthm}
\usepackage{epsfig}
\usepackage{epstopdf}
\usepackage{verbatim}
\usepackage{amsthm}
\usepackage{algorithmic}
\usepackage[linesnumbered,ruled,vlined]{algorithm2e}

\begin{document}

\title{Minimizing Outage Probability by Exploiting CSI in Wireless Powered Cooperative Networks}

\author{\IEEEauthorblockN{M. Majid Butt\IEEEauthorrefmark{1},~Ioannis Krikidis\IEEEauthorrefmark{2},~Nicola Marchetti\IEEEauthorrefmark{1}}
\IEEEauthorblockA{\IEEEauthorrefmark{1}CONNECT, Trinity College, University of Dublin, Ireland\\
\IEEEauthorblockA{\IEEEauthorrefmark{2}Department of Electrical and Computer Engineering, University of Cyprus\\
Email: \{buttm,~marchetn\}@tcd.ie, krikidis.ioannis@ucy.ac.cy}
}
}

%

\maketitle

\begin{abstract}
In this work, we address the relay selection problem for the wireless powered communication networks, where the relays harvest energy from the source radio frequency signals. A single source-destination pair is considered without a direct link.
The connecting relay nodes are equipped with storage batteries of infinite size.
We assume that the channel state information (CSI) on the source-relay link is available at the relay nodes. Depending on the availability of the CSI on the relay-destination link at the relay node, we propose different relay selection schemes and evaluate the outage probability. The availability of the CSI at the relay node on the relay-destination link considerably improves the performance due to additional flexibility in the relay selection mechanism. We numerically quantify the performance for the proposed schemes and compare the outage probability for fixed and equal number of wireless powered forwarding relays.

\end{abstract}

\section{Introduction}
Wireless powered communication networks (WPCNs) are one of the promising technologies to achieve sustainable wireless networks, where the communicating nodes are powered by radio frequency (RF) signals. The simultaneous wireless information and power transfer (SWIPT) concept has been investigated extensively to realize WPCNs. In SWIPT, energy and information are transferred from the same RF signal by using either time sharing or power splitting protocol \cite{Zhang_IEEECOM:2013,Ali_IEEEWS:2013}. The time sharing protocol allocates dedicated time for energy harvesting and information transfer, while power splitting extracts energy and information from the same RF signal.

Relays are used in wireless networks to extend the coverage and increase the information reliability. Relay selection problem using amplify-and-forward (AF) and decode-and-forward (DF) techniques in WPCNs has been addressed in literature quite extensively, e.g., \cite{Krikidis_COML:2012,Michalopoulos:JSAC2015}. A recent work in \cite{Michalopoulos:JSAC2015} discusses an RF based cooperative network, where the relays are used for transmitting information to a designated receiver and for transmitting energy to an associated ambient RF energy harvester. For the case where the number of relays is more than two, two relay selection methods are developed and the trade-off between outage probability and average energy transfer is discussed. The authors in \cite{Zhong:TCOM2014} analyze the performance of a network consisting of a single source, single relay and single destination. The throughput of the time sharing protocol for both AF and DF relaying schemes for the same model is also investigated in \cite{Nasir_TCOM:2015}. A similar system model is considered in \cite{Zhu_TCOM_2015}, where the relay is equipped with multiple antennas and the outage probability and ergodic capacity of the system are studied. Multiple source-destination pairs communicating through a single energy harvesting relay are considered in \cite{Ding_TWCOM:2013} and the system effect of the harvested energy distribution among the users is investigated.
\subsection{Motivation and Contributions}
In this work, we aim to minimize the outage probability for a cooperative system, comprising of a single source, multiple energy harvesting (EH) relays and a single destination.
A single relay is selected to forward the source signal to the destination.
We assume half duplex relay communication such that the relays receive the source signal in a time slot $t$ and the selected relay forwards it to the destination in time slot $t+1$.
The channels on both, source-relay and the relay-destination, links are mutually independent, and are independently and identically distributed (i.i.d). Due to mutual independence and i.i.d channel assumption on both links, relay selection poses new challenges as the relay selected to receive data from the source in time slot $t$ will have a completely independent (and unknown) channel realization in time slot $t+1$ for transmission on the relay-destination link.

For the similar setting, the work in \cite{majid:ietsp16} assumes that the CSI is not available on the source-relay link at the relay node. In contrast, we assume availability of the CSI on the source-relay link at the relay node throughout this work. When the CSI is available on the source-relay link, the relay selection exploits the CSI to decide which relays are dedicated for data/energy transfer. Then, conditioned on the availability of transmit CSI (CSIT) at relay on the relay-destination link, we propose novel relay selection schemes. The consideration of the availability of both transmit and receive CSI at the relay node requires a different relay selection approach as compared to relay selection in \cite{majid:ietsp16}.
We formulate the outage minimization optimization problem and evaluate the performance of the proposed relay selection schemes numerically. Then, we compare our schemes with the existing schemes available in literature and show their superiority in terms of outage performance.

This paper is structured as follows. Section \ref{sect:system} introduces the system model and the fundamentals for the problem. The novel schemes are proposed in Section \ref{sect:schemes} and the performance is evaluated numerically in Section \ref{sect:results}. Finally, Section \ref{sect:conclusion} concludes with the main contributions of the work.

\section{System Model}
\label{sect:system}
We consider a  system consisting of a single source S, a destination D and $N$ cooperating relays. A relay node $i$ is denoted by $L_i$. The source and the destination have no direct link and communicate through relays. We consider a broadcast channel between the source and the relays. Let us denote the source-relay and the relay-destination links by $S\to L_i$ and $L_i\to D$, respectively. The CSIT is not available at the source and therefore, the source transmits with a fixed power $P_s$.
The received signal $ y_i(t)$ at the relay $L_i$ is expressed as,
\begin{equation}
y_i(t)= \frac{1}{\sqrt{d_i^{\alpha}}}\sqrt{P_s}h_{si} x(t)+n(t),
\label{3}
\end{equation}
where $ x(t) $ and $d_i$ denote the normalized information signal from the source and the distance between the transmitter and the relay $L_i$, respectively. Without loss of generality, $d$ is assumed to be unity throughout the work to focus on the main results. $\alpha$ represents the path loss exponent. $n(t)\sim Z(0,\sigma^2)$ is the Gaussian noise with zero mean and variance $\sigma^2$, while small scale fading on the $S\to L_i$ link is represented by $h_{si}$.

\begin{figure}
\centering
  	\includegraphics[width=3.0in]{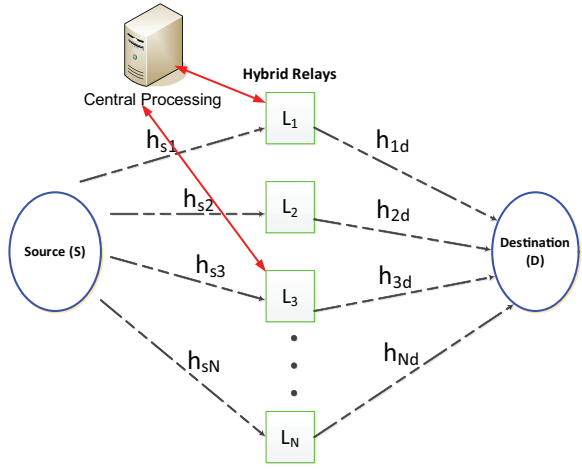}
   \caption{Schematic diagram for the system model. The centralized controller collects relay stored energy information and CSI at relay on the $S\to L_i$ link for all the relays and makes relay selection. Note that all the links from the relays to the centralized server are not shown to make the diagram clear.}
	\label{fig:system}
\end{figure}

We make the following assumptions regarding the cooperative system.
\begin{itemize}
\item We consider block fading i.i.d channels on both $S\to L_i$ and $L_i\to D$ links such that the channels remain constant for the duration of a time slot, but vary independently from slot to slot. The schematic diagram for the system model is shown in Fig. \ref{fig:system} with a centralized controller, which collects stored energy and channel state information from all the relays and makes relay selection in every time slot.
\item We use time sharing protocol for SWIPT at relay nodes.
\item The transmissions are interference free and interference cannot be exploited for energy harvesting at relays.
\item We consider a half duplex communication system. The selected relay node $L_{i^*}$ decodes information from the source in a time slot of duration $T$ and forwards it to the destination in the next time slot. Hence, node $L_{i^*}$ is not available for information (or energy) reception in time slot $t+1$. However, all other nodes are available for information/energy reception from the source signals, thereby mimicking a full duplex communication system \cite{Ikhlef_TVT:2012}.
\item The circuit energy consumed in energy harvesting and information decoding at relay is negligible.
\item A single relay is selected for forwarding information to the destination. It is well known that transmission from multiple relays provides spatial diversity and improves data reliability at the cost of increased complexity. We focus on single relay selection schemes to reduce the complexity of the system, but all the proposed schemes can be extended to multiple relay transmission schemes in a straight forward manner.
\item The energy is stored in a battery of infinite capacity and it is assumed that there is negligible leakage within the time period of interest \cite{Yuan:ICC2014}.
\end{itemize}
The rate $R_{si}$ provided on the $S\to L_i$ link in a time slot $t$ is given by
\begin{equation}
R_{si}= \frac{1}{2}\log_2 \big(1+|h_{si}|^{2}\frac{P_s}{\sigma^2}\big).
\label{eqn:rate_equation1}
\end{equation}
For a relay transmit power $P_{r}$, the rate $R_{id}$ provided by the $L_i\to D$ link is given by,
\begin{equation}
R_{id}= \frac{1}{2}\log_2 \big(1+|h_{id}|^{2}\frac{P_{r}}{\sigma^2}\big)~,
\label{eqn:rate_equation2}
\end{equation}
where $h_{id}$ denotes the channel coefficient between the relay node $L_i$ and the destination D.

All the nodes selected for energy transfer harvest energy from the source signals. Assuming $T=1$ without loss of generality, the energy harvested by a relay node is given by
\begin{equation}
E_i^h=\eta P_s\left | h_{si} \right |^2,
\label{4}
\end{equation}
where $0\leq\eta\leq 1$ is the energy harvesting efficiency.

\subsection{Problem Formulation}
For a traditional grid powered DF relaying strategy, the outage probability $P_{\rm out}$ that a rate $R$ is not supported by the system is given by
\begin{eqnarray}
P_{\rm out}&=&\Pr\Big(\min(R_{si^*},R_{i^*d})<R\Big)\\
&=&{\Pr} \Big\{\min\Big(\frac{1}{2}\log_2 (1+ | h_{si^*}|^{2}\frac{P_s}{\sigma^2}),\nonumber\\
&&\frac{1}{2}\log_2 (1+ | h_{i^*d}|^{2}\frac{P_{r}}{\sigma^2})\Big)<R \Big\},
\label{eqn:DF}
\end{eqnarray}
where $L_{i^*}$ is the selected node for information relaying.
However, when the relays are powered by EH, unavailability of the harvested energy is an additional source of outage.

For a wireless powered relay network, the outage probability is given by,
\begin{equation}
P_{\rm out}= 1-\zeta_s\zeta_p\zeta_r,
\label{eqn:outage_EH}
\end{equation}
where $\zeta_s$ and $\zeta_r$ denote the success probability on the $S\to L_i$ and $L_i\to D$ links, respectively while $\zeta_p$ is the probability that the selected relay can support a transmit power $P_r$ to forward the information to the destination. $\zeta_p$ depends on the relay selection scheme and energy harvesting efficiency $\eta$. If there is always sufficient energy available for decoding, $\zeta_p\to 1$ and the EH system behaves as a grid powered system.

Our objective is to find a relay selection policy $\pi$ that minimizes the outage probability for the EH cooperative system with a fixed number of relays.
The optimization problem is formulated as:
\begin{eqnarray}
&\min_{\pi}&~ {P_{\rm out}}\\
&{\rm s.t}.&~
\begin{cases}
\mathcal{C}_1:N=c_1,& c_1\in \mathbb{N}\\
\mathcal{C}_2:P_s = c_2\\
\mathcal{C}_3:R\geq 0\\
\mathcal{C}_4:R_{si}=\frac{1}{2}\log_2 (1+ | h_{si}|^{2}\frac{P_s}{\sigma^2})& i\leq N\\
\mathcal{C}_5:R_{id}=\frac{1}{2}\log_2 (1+ | h_{id}|^{2}\frac{P_{r}}{\sigma^2}) & i\leq N\\
\mathcal{C}_6:E_i^h=\eta P_s\left | h_{si} \right |^2& i\leq N\\
\mathcal{C}_7:E_{i^*}^{\rm st}>E_r
\end{cases}
\label{eqn:cons}
\end{eqnarray}
where $c_1$ and $c_2$ are constants representing a fixed number of relay nodes and the fixed source power, respectively. $\mathcal{C}_3-\mathcal{C}_6$ represent the rate and energy harvesting conditions according to the channel distributions on the $S\to L_i$ and $L_i\to D$ links. $\mathcal{C}_7$ is neutrality constraint which implies that stored energy $E_{i^*}^{\rm st}$, for the node selected for forwarding $L_{i^*}$, must be greater than the energy $E_r$ required for transmission. We intend to find a relay selection policy $\pi$ which follows the constraints in ($\ref{eqn:cons}$) and minimizes network failure (outage probability).

A closed-form solution for the problem is difficult to achieve due to involvement of multiple energy queues at the relay nodes, which depend on stochastic fading channels. The energy queue states are mutually coupled for the outage analysis, making computation of $\zeta_p$ in (\ref{eqn:outage_EH}) difficult, and the analysis is not tractable for large number of relay energy queues. Therefore, we propose heuristic relay selection schemes and evaluate the outage performance numerically.

\section{Relay Selection Schemes}
\label{sect:schemes}
To avoid the outage event, there must be at least a single relay available with sufficient energy to transfer data to the destination in time slot $t+1$. To minimize the outage probability, the relay selection scheme should aim at maximizing energy harvesting of the relays and minimizing the energy expenditure on the $L_i\to D$ link, thereby maximizing the network lifetime with minimum failure.

A simplified approach to model such a system is to assume the length of fading blocks long enough such that the channels on the $S\to L_i$ and $L_i \to D$ links remain constant for both reception and forwarding phases at the relay \cite{Yaming,Ju_IEEEWCOM:2014}. This has the advantage that both receive and transmit channels are known at the time of relay selection. In contrast, we assume that the relay reception and transmission occurs in two consecutive time slots.
Due to mutually independent channels on the $S\to L_i$ and $L_i\to D$ links, and the fact that the CSI for the $L_i \to D$ channel is not available when a relay is selected for forwarding at time $t$, relay selection becomes challenging.

The relay selection scheme is dictated by the availability of the CSI at relay on the both $S\to L_i$ and $L_i \to D$ links. Throughout this work, we assume that the CSI is available at the relay on the $S\to L_i$ link\footnote{The CSI estimation at the relay can be performed by pilot/data aided techniques.}.

Regarding the availability of the CSI at relay node on the $L_i \to D$ link, we consider the following two scenarios:
\begin{enumerate}
  \item The CSIT is not available on the $L_i \to D$ link.
  \item The CSIT is available on the $L_i \to D$ link.
\end{enumerate}

\subsection{No CSIT on the $L_i \to D$ Link}
When the CSIT is not available on the $L_i \to D$ link, no power allocation can be performed. Therefore, the selected relay transmits with a fixed power $P_{r}$ at time $t+1$.

At time $t$, the relay selection is performed. As the CSIT on $L_i \to D$ link is not available, relay selection is solely based on the available information at time $t$. As a single relay is selected at time $t$, this scheme is called single relay selection with No CSIT (SRS-NCSI).

A relay is selected for decoding information such that,
\begin{eqnarray}
\label{eqn:SRS_selection}
i^*=\arg \min_i R_{si}
\end{eqnarray}
where (\ref{eqn:SRS_selection}) is evaluated for a relay $i$ only if,
\begin{equation}\label{eqn:SRS}
   I(R_{si}>R)\times I\big(\frac{E_i^{\rm st}}{T}>P_r\big)=1
\end{equation}
such that,
\begin{equation}
I(R_{si}>R)=\begin{cases}0 & R_{si}<R\\
1 & R_{si}\geq R
\end{cases},
\label{eqn:SRS1}
\end{equation}
and
\begin{equation}
I\Big(\frac{E_i^{\rm st}}{T}>P_r\Big)=\begin{cases}0 & \frac{E_i^{\rm st}}{T}<P_r\\
1&\frac{E_i^{\rm st}}{T} \geq P_r
\end{cases}.
\label{eqn:SRS2}
\end{equation}
$E_i^{\rm st}$ denotes the stored energy for relay $L_i$.
The indicator functions $I(R_{si}>R)$ and $I\big(\frac{E_i^{\rm st}}{T}>P_r\big)$ in (\ref{eqn:SRS}) ensure that a selected node can decode the signal from the source and has energy to transmit with a fixed power $P_{r}$ in time slot $t+1$. Equation (\ref{eqn:SRS_selection}) selects the node with the minimum $R_{si}$ for information decoding out of the nodes which satisfy (\ref{eqn:SRS}).
The rationale behind the selection of the node with minimum $R_{si}$ is to provide relatively 'average' $S\to L_i$ channel for information decoding at relay as information decoding is already ensured by the condition $I(R_{si}>R)$. This implies that good $S\to L_i$ channels (which satisfy (\ref{eqn:SRS})) can be better utilized for energy harvesting as decoding on the best channel does not improve the outage performance as long as (\ref{eqn:SRS1}) is satisfied. If $R_{i^*d}<R$ for the selected relay or no relay satisfies (\ref{eqn:SRS}), an outage event occurs.

All the relays with $i\ne i^*$ harvest energy from the source signal such that,
\begin{equation}
E_i^{\rm st}(t+1) = E_i^h(t)+ E_i^{\rm st}(t), \quad i\ne i^*~.
\end{equation}
The selected relay node $L_{i^*}$ is not a candidate for selection for both decoding and harvesting from the source signal in time slot $t+1$, which implies that $E_{i^*}^{\rm st}(t+1)=E_{i^*}^{\rm st}(t)$ and the stored energy for node $L_{i^*}$ after making a transmission at time $t+1$ is given by,
\begin{equation}
E_{i^*}^{\rm st}(t+2)= E_{i^*}^{\rm st}(t+1)-P_{r}T~.
\end{equation}
It is clear from (\ref{eqn:DF}) that $P_{\rm out}$ is determined by the rate provided by the 'bottleneck' link. However, the outage probability in WPCN is also characterized by the amount of energy harvested by the relay nodes. The harvested energy is a function of the source power, channel distribution and the energy harvesting efficiency $\eta$. When $\eta$ is large, very small number of relay nodes provide enough stored energy such that there is always a node available with enough energy to transmit on the $L_i \to D$  link and the outage probability in (\ref{eqn:outage_EH}) converges to (\ref{eqn:DF}). However, when $\eta$ or $N$ is small, (\ref{eqn:DF}) is only a lower bound on $P_{\rm out}$.


\subsection{CSIT Available on the $L_i \to D$  Link}
\label{sect:scheme_ACSI}
In the case when the CSIT is available at the relay on the $L_i \to D$ link, the relay can benefit from this information through power allocation. The required transmit power to make a successful transmission for a relay $L_i$ is computed from (\ref{eqn:rate_equation2}), and given by,
\begin{equation}
P_{id} = \frac{(2^{2R}-1)\sigma^2}{|h_{id}|^{2}}.
\label{eqn:power}
\end{equation}
This scenario provides more flexibility for relay selection. However, the CSIT on the $L_i \to D$ link is available only at time $t+1$ due to i.i.d channel assumption and the relay is selected at time $t$.

The main challenges in relay selection are:
\begin{enumerate}
  \item If the relay selection is made based on the channel quality on the $S\to L_i$ link, the selected relay node may not have enough energy to transmit on the $L_i \to D$ link at time $t+1$. At the same time, the use of channel quality on the $L_i \to D$ channel will not be optimal as the selected relay $L_{i^*}$ may not necessarily have the best channel at time $t+1$.
  \item If the relay selection is based on the stored energy maximization, the availability of the CSI on $S\to L_i$ and $L_i \to D$ links at the relay node is not exploited.
\end{enumerate}
To take the advantage of CSI availability at different times, we propose a two step relay selection algorithm.\\
\textbf{Phase I}: In the first phase, a subset $\Gamma$ of (maximum) $M$ relays\footnote{To avoid confusing it with multiple relay forwarding, please note that only one relay will be selected for forwarding after phase 2 of the scheme.} is selected out of $N$ relays for decoding information. As the CSIT on the $L_i \to D$ link is not known at time $t$, more than one relay decode information to provide multiuser diversity for the transmission on the $L_i \to D$ link. It is worth noting that selecting a single relay in phase 1 makes available CSIT on the $L_i \to D$ link at time $t+1$ useless as all other relays cannot be used for forwarding in phase 2 of the scheme. Due to multiple relay selection in first phase, this scheme is termed as Multiple Relay Selection with available CSI (MRS-ACSI).

The selection for the forwarding set $\Gamma$ is made such that,
\begin{eqnarray}
\label{eqn:csi_gamma2}
\Gamma^{K\times 1}=\{i:\gamma_i\leq \gamma_K\}
\end{eqnarray}
where (\ref{eqn:csi_gamma2}) is evaluated for a relay $i$ only if,
\begin{equation}
  I(R_{si}>R)=1.
\end{equation}
$\gamma_K$ denotes $K^{\rm th}$ smallest fading channel selected out of $U$ nodes satisfying $I(R_{si}>R)$. Cardinality $K$ of $\Gamma$ is limited by $\min(M,U)$, where $M\leq N$ is a system parameter for the scheme. Equation (\ref{eqn:csi_gamma2}) states that $K$ relays with the smallest fading channels are dedicated for decoding information.
This metric chooses the relay nodes with the weakest channels, but capable of decoding the information. This implies that the rest of the $N-K$ relays harvest energy from the source signals on good channels and the stored energy for the nodes increases at a faster rate.\\
\textbf{Phase II:} In the second phase of the relay selection algorithm, the forwarding relay from the set $\Gamma$ at time $t+1$ is selected such that,
\begin{eqnarray}
\label{eqn:MRS_Ph2}
i^*=\arg \min_{i\in \Gamma} P_{id}
\end{eqnarray}
where (\ref{eqn:MRS_Ph2}) is evaluated for a relay $i$ only if,
\begin{equation}
  I\Big(\frac{E_i^{\rm st}}{T}>P_{id}\Big)=1
\label{eqn:MRS_cons}
\end{equation}
such that,
\begin{equation}
I\Big(\frac{E_i^{\rm st}}{T}>P_{id}\Big)=\begin{cases}
1& \frac{E_i^{\rm st}}{T}\geq P_{id}\\
0&\frac{E_i^{\rm st}}{T}<P_{id}
\end{cases}.
\label{eqn:MRS_ph2cond}
\end{equation}
The scheme selects the relay with the best transmit channel out of the relays, which have enough stored energy for transmission as in constraint (\ref{eqn:MRS_cons}). This ensures transmission with minimum expenditure and is optimal decision for the relays in $\Gamma$. Note that $P_{id}$ is calculated individually for every relay $L_i$ via (\ref{eqn:power}). If the cardinality of $\Gamma$ set is zero or no relay in the set satisfies (\ref{eqn:MRS_cons}), outage occurs.
%

After transmission, the stored energy for the node $L_{i^*}$ is updated as,
\begin{eqnarray}
E_{i^*}^{\rm st}(t+2)= E_{i*}^{\rm st}(t+1)-P_{i^*d}(t+1)T~.
\end{eqnarray}
The rest of the nodes harvest and store energy depending on the received signal strength from the source such that
\begin{eqnarray}
E_j^{\rm st}(t+1)=
\begin{cases}
E_j^h(t)+ E_j^{\rm st}(t),&   j \notin \Gamma\\
E_j^{\rm st}(t),&   j \in \Gamma,j \ne i^*
\end{cases},
\end{eqnarray}
with the nodes $j\in \Gamma,j\ne i$ not able to harvest energy as they were reserved for decoding.

There is a tradeoff involved with the selection of parameter $M$ for a fixed $N$. If $M$ is large, there is greater chance of finding a good channel for transmission on the $L_i \to D$ link, but fewer relays are available for EH and the relay system becomes power limited. On the contrary, if $M$ is too small, less multiuser diversity is exploited on the $L_i \to D$  link, but more relays harvest energy. Thus, for the proposed scheme, it is important to optimize $M$ for a given $N$ and $\eta$.

Given that we have a multiple relay selection (MRS-ACSI) policy $\pi(M,N)$, the parameter optimization problem is formulated by
\begin{eqnarray}
M^*(N,\eta,R) &=& \arg\min_{\pi(M,N),~0<M\leq N} P_{\rm out},
\end{eqnarray}
with the same constraints as in (\ref{eqn:cons}). We determine the optimal $M$ for the proposed scheme numerically in Section \ref{sect:results}.

\section{Numerical Results}
\label{sect:results}
We numerically evaluate the performance of the proposed schemes in this section. A Rayleigh fading channel with mean one is considered on the $S\to L_i$ and $L_i \to D$ links. Time slot $T$ is assumed to be one while noise variance $\sigma^2=1$. $P_s$ is fixed to $10$ dBW throughout.

\begin{figure}
\centering
  	\includegraphics[width=3.5in]{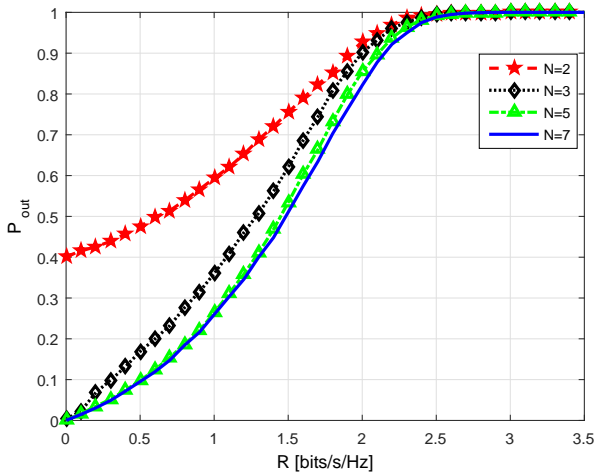}
   \caption{Outage probability for the SRS scheme for different $N$ and $\eta=0.7$.}
	\label{fig:SRS}
\end{figure}

In Fig. \ref{fig:SRS}, we compute the outage probability for the SRS-NCSI scheme for different number of available relay nodes. The CSIT is not available at the relay node on the $L_i \to D$ link and the selected relay node transmits with a fixed power 10 dBW. The outage probability decreases as the number of relays increases, as expected. However, when $N$ is sufficiently large, any further increase in $N$ does not benefit. For a small $N$, there is high probability that the selected relay stored energy $E_i^{\rm st}$ is not enough to transmit successfully on the $L_i \to D$ channel and 'power limitation' of the relay node contributes to the outage significantly. As $N$ increases, the outage performance improves. At $N=7$, the effect of power limitation vanishes completely and $N>7$ does not help to decrease outage. The system behaves like a grid powered system and the outage performance is given by (\ref{eqn:DF}).

We compare SRS-NCSI scheme with other similar available schemes. As a benchmark, we consider two commonly used schemes. In the first scheme, the relay is selected such that \cite{majid:ietsp16,Yaming,Krikidis_COML:2012},
\begin{equation}
i^* = \arg\max_i (E_i^{\rm st}-P_r)^+~\times I(R_{si}>R),
\label{eqn:bestenergy}
\end{equation}
with the notation $x^+=\max(x,0)$.
We denote it by SRS-NCSI-best-energy scheme, where the relay with the largest residual energy is selected for transmission.

\begin{figure}
\centering
  	\includegraphics[width=3.5in]{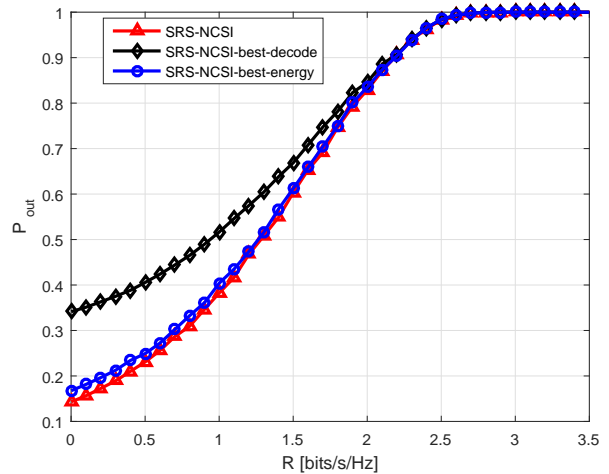}
   \caption{Comparison of the relay selection schemes for the case when $N=10$ and $\eta=0.1$, and the CSIT is not available at relay on the $L_i \to D$  link.}
	\label{fig:SRS_comp}
\end{figure}

The second scheme selects the relay which has the best chance of decoding on the $S\to L_i$ channel. The concept is similar to relay antenna selection scheme in \cite{Krikidis_TCOM2:2014}, where the antennas with large channel gains are selected for decoding. Thus,
\begin{equation}
i^* = \arg\max_i R_{si}\times I(R_{si}>R)~.
\label{eqn:bestdecoding}
\end{equation}
We denote this scheme by SRS-NCSI-best-decoding. Please note that the forwarding in all schemes is made only if $E_i^{\rm st}>P_r$, which saves transmit energy on unsuccessful transmission.

From Fig. \ref{fig:SRS_comp}, we see that SRS-NCSI outperforms the other schemes. The SRS-NCSI-best-energy performs better than SRS-NCSI-best-decoding because the major cause of outage is insufficient energy to forward data for the selected relay. The best channel selection on the $L_i \to D$ link is not optimal as decoding is already ensured by the condition $I(R_{si}>R)$.

\begin{figure}
\centering
  	\includegraphics[width=3.5in]{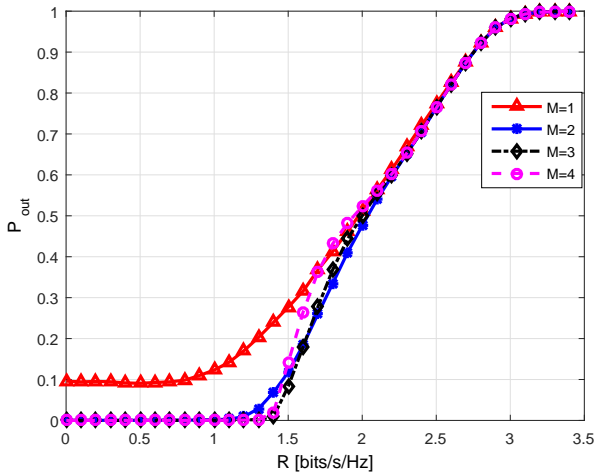}
   \caption{Outage probability for the MRS scheme for $N=10$ and $\eta=0.1$.}
	\label{fig:MRS}
\end{figure}

In Fig. \ref{fig:MRS}, we plot the outage probability for the MRS-ACSI scheme and compute the optimal value of parameter $M$ for a given $N$ and $\eta$. If $M$ is too small, multiuser diversity is not exploited effectively on the $L_i \to D$ link. On the contrary, if $M$ is too large, EH is not enough for the relays to store enough energy to avoid outage events. We observe that $M=3$ is optimal at small $R$, while $M=2$ is optimal at large $R$. This is attributed to the fact that large rate requirements require more relay nodes to harvest energy to have sufficient energy for successful transmissions to the destination.

\begin{figure}
\centering
  	\includegraphics[width=3.5in]{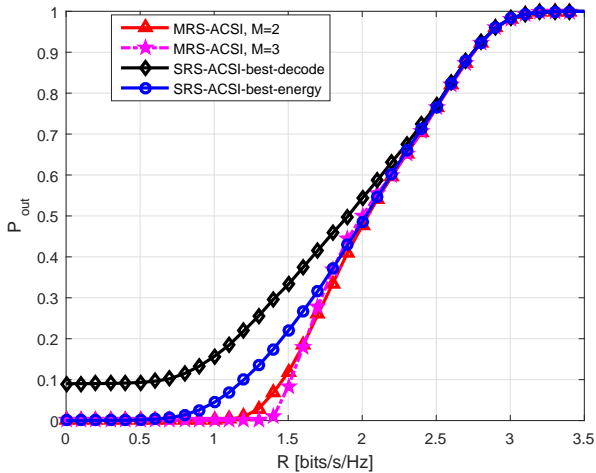}
   \caption{Comparison of the relay selection schemes for the case when $N=10$ and $\eta=0.1$, and the CSIT is available at the relay on the $L_i \to D$  link.}
	\label{fig:MRS_comp}
\end{figure}

In Fig. \ref{fig:MRS_comp}, we compare the performance of the MRS scheme with the two schemes mentioned above. We assume that the CSIT is available at relay on the $L_i \to D$ link and power $P_{id}$ is allocated for the selected relay by (\ref{eqn:power}). However, due to unavailability of $P_{id}(t+1)$ at time $t$, we eliminate transmit power term from (\ref{eqn:bestenergy}) and evaluate the metric,
\begin{equation}
i^* = \arg\max_i E_i^{\rm st}(t)~\times I(R_{si}>R)~.
\end{equation}
It is worth noting that the relay $L_{i^*}$ is not available for harvesting at time slot $t$ even if $E_i^{\rm st}<P_{id}(t+1)$ because $P_{id}$ can only be calculated at instant $t+1$ due to delayed CSIT on the $L_i \to D$ link. Fig. \ref{fig:MRS_comp} shows that power allocation due to available CSIT on the $L_i \to D$ channel at time $t+1$ improves the outage performance for the SRS-NCSI-best-energy and SRS-NCSI-best-decoding schemes as compared to their respective performance in Fig. \ref{fig:SRS_comp}, but MRS-ACSI outperforms both schemes comfortably due to inherent multiuser diversity exploitation.

\begin{figure}
\centering
  	\includegraphics[width=3.5in]{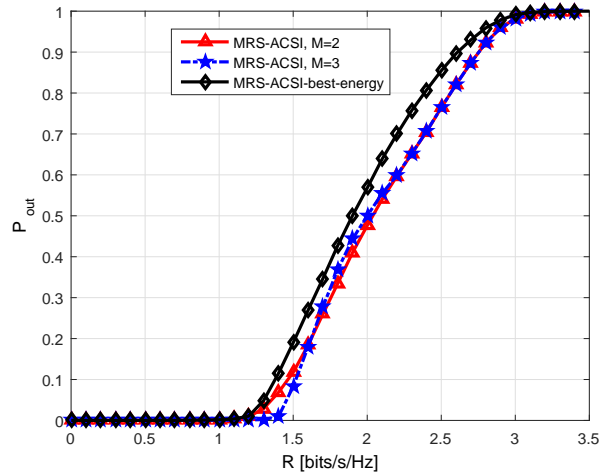}
   \caption{Comparison of the relay selection schemes for the case when $N=10$ and $\eta=0.1$, and the CSIT is available at the relay on the $L_i \to D$  link. Multiuser diversity is exploited using different relay selection metrics.}
	\label{fig:MRS_comp_group}
\end{figure}

To demonstrate the effect of relay selection metrics in (\ref{eqn:csi_gamma2}) and (\ref{eqn:MRS_Ph2}), we compare MRS-ACSI scheme with a similar 2-phase relay selection scheme proposed in \cite{majid:ietsp16}. Like MRS-ACSI, $M$ relays with the largest stored energies are selected in first phase. In the second phase, a relay $i$ out of $M$ relays is selected such that,
\begin{equation}
i^* = \arg\max_{i\in \Gamma} \Big(E_{\rm st}^i(t+1)-P_{id}(t+1)T\Big).
\end{equation}
This scheme is denoted by MRS-ACSI-best-energy. For the MRS-ACSI-best-energy scheme, $M^*=5$ for the parameters $\eta=0.1, N=10$ \cite{majid:ietsp16}. The results in Fig. \ref{fig:MRS_comp_group} reveal that our scheme performs better that the MRS-ACSI-best-energy scheme in spite of the fact that both schemes are exploiting multiuser diversity. From the numerical evaluation in Fig. \ref{fig:MRS_comp} and Fig. \ref{fig:MRS_comp_group}, we conclude that based on exploitation of multiuser diversity and careful design metrics for both phases as explained in Section \ref{sect:scheme_ACSI}, our proposed MRS-ACSI scheme performs better that the other schemes available in literature.

\section{Conclusions}
\label{sect:conclusion}
We propose novel relay selection schemes for the WPCNs and discuss the scenarios where the CSI is available at relay on the $S\to L_i$ link. Conditioned on the availability of the CSIT on the $L_i \to D$ link, we propose various relay selection schemes. We evaluate the performance of the proposed schemes numerically and compare it with the commonly used relay selection schemes. When the CSI is available at the relay on both $S\to L_i$ and $L_i \to D$ links, consideration of mutually independent i.i.d channels on two hops makes the half duplex relay selection problem challenging. A two--phase relay selection scheme in conjunction with our proposed relay selection metrics is proposed to exploit the multiuser diversity effectively. The numerical evaluation shows that our proposed scheme outperforms the other schemes comfortably when power allocation is applied on the $L_i \to D$ link. 

\section*{Acknowledgement}
This publication has emanated from research supported in part by a research grant from Science Foundation Ireland (SFI) and is co-funded under the European Regional Development Fund under Grant Number 13/RC/2077. Part of this work was supported by the Research Promotion Foundation, Cyprus,
under the Project FUPLEX with Pr. No. CY-IL/0114/02.

\bibliographystyle{IEEEtran}
\bibliography{bibliography}
\end{document}